\documentclass[a4paper]{jpconf}
\usepackage{graphicx,amssymb,amsmath}

\newcommand{\beq}{\begin{equation}}
\newcommand{\eeq}{\end{equation}}
\newcommand{\ben}{\begin{eqnarray}}
\newcommand{\een}{\end{eqnarray}}

\newcommand{\ie}{\mbox{\it i.e.}}
\newcommand{\eg}{\mbox{\it e.g.}}

\newcommand{\mchi}{\mbox{$m_\chi$}}
\newcommand{\sigv}{\mbox{$\langle \sigma v \rangle$}}

\begin{document}

\title{Cosmic-ray antiproton constraints on light dark matter candidates}

\author{Julien Lavalle$^{\star,\dagger,}$\footnote{former Multidark fellow}}

\address{$^\star$ (perm. addr.) Lab. Univers et Particules de Montpellier (LUPM), CNRS-IN2P3 \& Univ. Montpellier II 
  (UMR 5299), Place Eug\`ene Bataillon, F-34095 Montpellier cedex 05 --- France}

\address{$^\dagger$ Instituto de F\'isica Te\'orica (UAM/CSIC) \& Dep$^\text{to}$ de F\'isica Te\'orica (UAM),
  Univ. Aut\'onoma de Madrid, Cantoblanco, E-28049 Madrid --- Spain}

\ead{lavalle@in2p3.fr}

\begin{abstract}
Some direct detection experiments have recently collected excess events that could be interpreted as a dark 
matter (DM) signal, pointing to particles in the $\sim$10 GeV mass range. We show that scenarios in which 
DM can self-annihilate with significant couplings to quarks are likely excluded by the cosmic-ray (CR) 
antiproton data, provided the annihilation is S-wave dominated when DM decouples in the early universe. These
limits apply to most of supersymmetric candidates, \eg~in the minimal supersymmetric standard model (MSSM) and
in the next-to-MSSM (NMSSM), and more generally to any thermal DM particle with hadronizing annihilation final states.
\end{abstract}

\section{Introduction}
\label{sec:intro}
After the finding of an annual modulation consistent with a DM signal in the DAMA/Libra data 
\cite{2000PhLB..480...23B,2008EPJC..tmp..167B}, some skepticism had arisen due to the failure of other direct 
detection experiments to confirm it in the preferred phase space corner. Interestingly enough, 
this initial caution has been recently tickled by the results of a few other experiments with different setups. 
CoGeNT \cite{2011PhRvL.106m1301A} and CRESST \cite{2011arXiv1109.0702A} have indeed found excess events; CoGeNT 
has even released hints for an annual modulation \cite{2011PhRvL.107n1301A}. Yet, XENON-10/100 
\cite{2011PhRvL.107e1301A,2011PhRvL.107m1302A}, CDMS \cite{2011PhRvL.106m1302A} and SIMPLE \cite{2011arXiv1106.3014F} 
do not validate such excesses and therefore challenge the classical DM interpretation of the aforementioned positive 
results --- \ie~the elastic collisions of weakly massive interacting particles (WIMPs) off target nuclei
\cite{2011arXiv1110.2721K}. It is still important to recall that these excesses --- and bounds --- rely on data 
analyses led very close to the current experimental thresholds, where systematic errors can in principle get large 
and remain to be much better controlled. Nevertheless, these measurements have brought some excitement in the DM 
community, and many authors have tried to fit the available data with some phenomenological models 
(\eg~\cite{2010PhRvD..81j7302B}). The preferred mass range is around 10 GeV, while difficulties are generally found 
when trying to get good fits to the signals while avoiding the constraints from other direct detection experiments 
\cite{2011arXiv1110.2721K}. Some ideas have been proposed to explain such discrepancies 
(\eg~\cite{2011JCAP...09..022A,2011PhRvD..84d1301F}), but what people usually do is to forget about either the 
bounds or the signals, depending on whether they want to promote their models or challenge them.

Beside direct detection, indirect detection, that looks for DM annihilation or decay traces originating in 
high concentration astrophysical sites, provides an interesting way to check the relevance of any 
phenomenological proposal in this context. Among indirect signatures, CR antiprotons have long been 
known as an interesting target to look at~\cite{1984PhRvL..53..624S}. In this proceedings' 
contribution, we show that they indeed provide quite strong limits on quarkophilic DM candidates, and are 
even likely excluding those models which have an S-wave dominated annihilation cross section. The 
following discussion is mostly based upon two articles, Refs. \cite{2010PhRvD..82h1302L} and 
\cite{2012NuPhB.854..738C}, to which we refer the reader for more details and references.
\section{Ingredients and results}%
\begin{figure}[t!]
\centering
\includegraphics[width=0.32\textwidth]{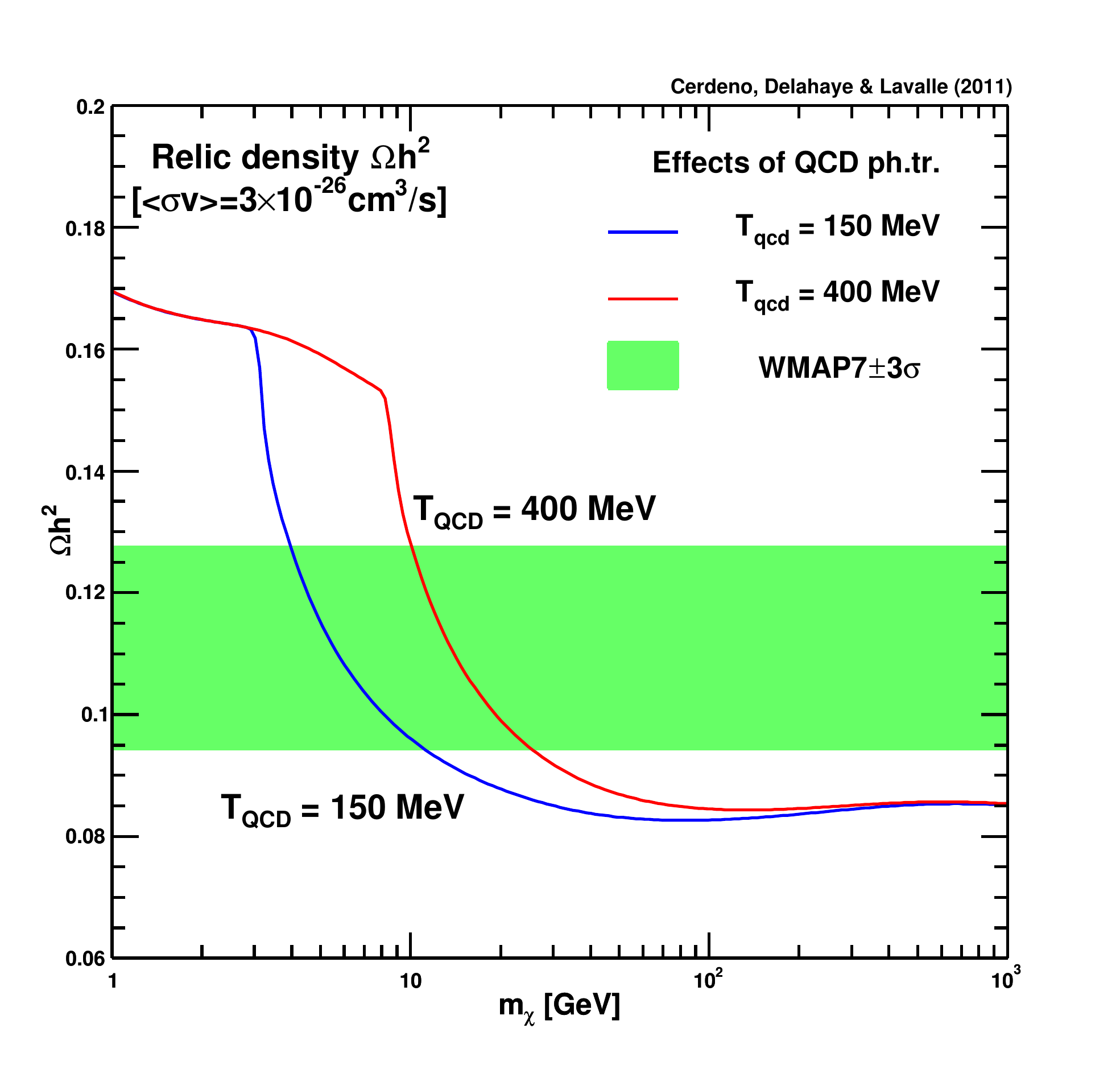}
\includegraphics[width=0.32\textwidth]{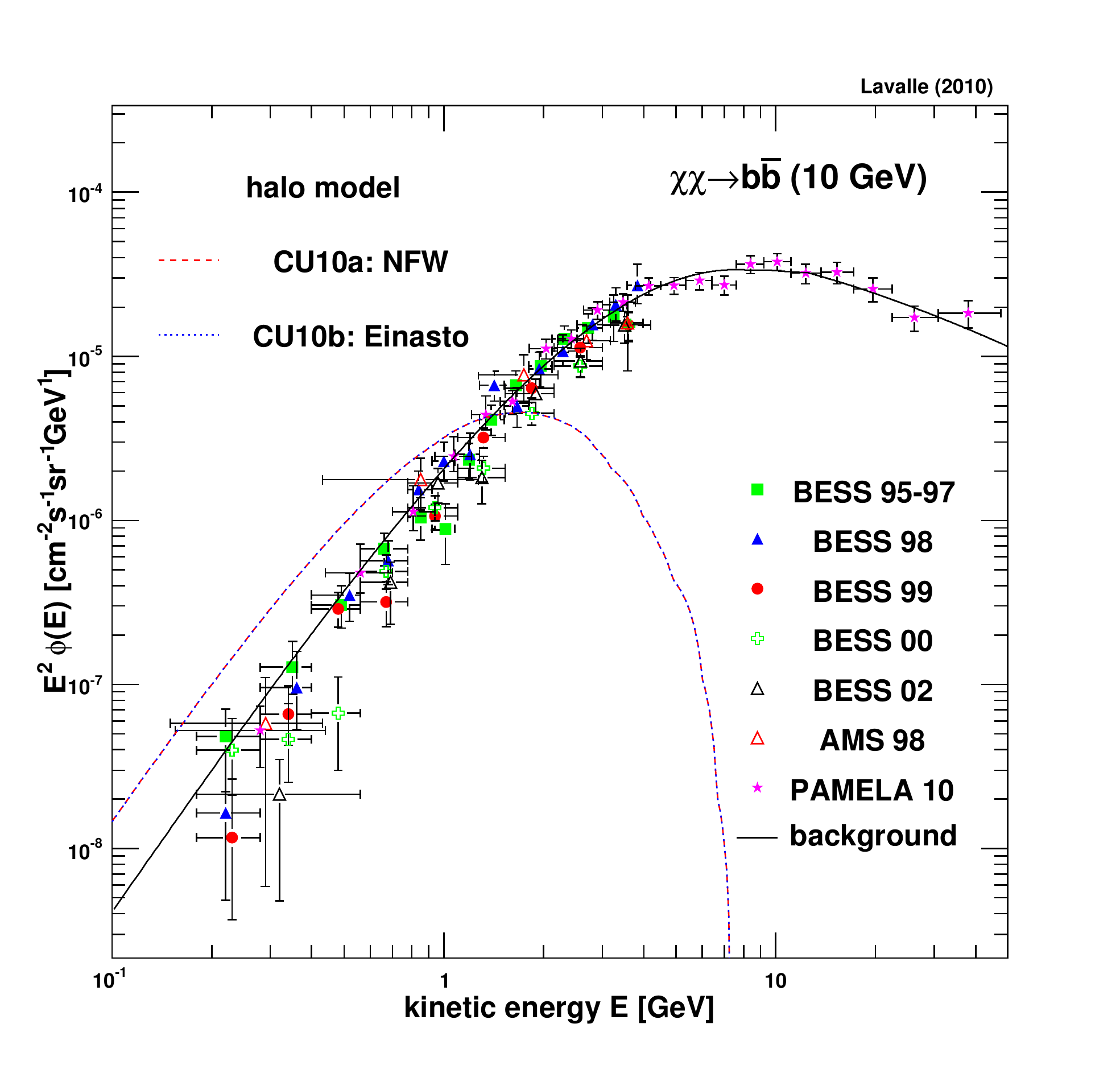}
\includegraphics[width=0.32\textwidth]{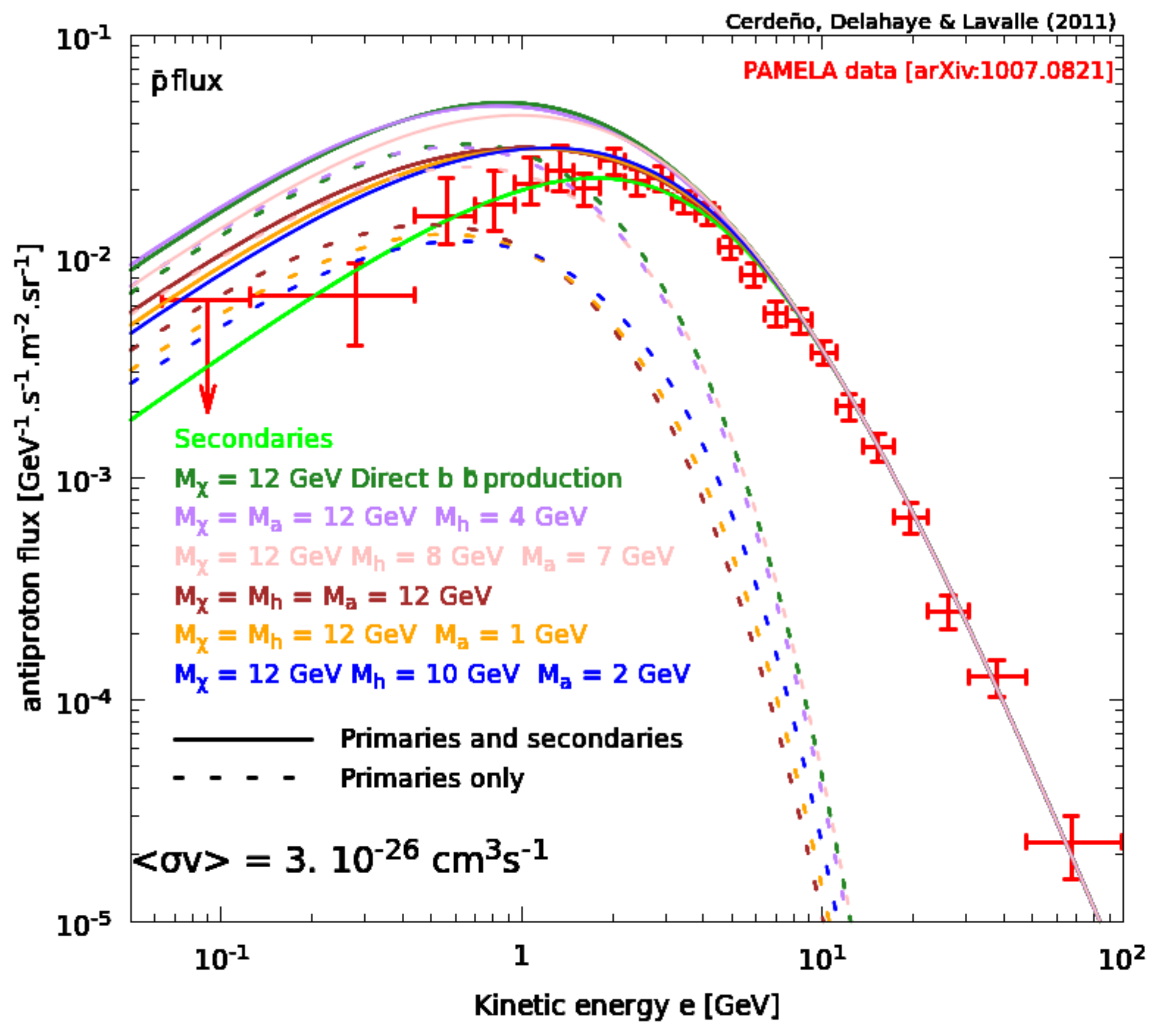}
\caption{Left: Impact of considering 2 different temperatures for the quark-hadron transition on the relic density
calculation as a function of the WIMP mass, for a given annihilation cross section \cite{2012NuPhB.854..738C}.
Middle: antiproton flux prediction for a 10 GeV WIMP annihilating into $b\bar b$ \cite{2010PhRvD..82h1302L}. Right: 
antiproton flux predictions for a 12 GeV WIMP annihilating into different mass combinations of an intermediate 
two-boson state which further decays into quarks \cite{2012NuPhB.854..738C}.}
\label{fig:lavalle_figs}
\end{figure}

\label{sec:ingredients}%
Before discussing the Galactic ingredients, it might be useful to recall a source of uncertainty in the 
annihilation cross section that comes from the calculation of the relic density, though commonly assumed to be 
under control. For typical thermal cold DM candidates, the chemical decoupling that sets the DM abundance occurs
in the early universe at a temperature $T_f \approx \mchi / 20 $. In the 10 GeV mass range, this translates into
temperatures around 500 MeV, a temperature range in which the quark-hadron transition is expected 
\cite{1988NuPhB.310..693S,1991NuPhB.360..145G}. A decoupling occurring before or after this transition
induces a factor of 2 uncertainty in the annihilation cross section for a given cosmological abundance, since the 
relativistic degrees of freedom, which characterize the energy density in the radiation era (\ie~the squared 
expansion rate), vary by a factor of 4 meanwhile. We therefore note as a preliminary remark that the required 
annihilation cross section may be twice larger than the canonical value in this mass range, the transition
mass being as uncertain as the quark-hadron transition temperature. This was discussed in detail in 
\cite{2012NuPhB.854..738C}, and is illustrated in the left panel of Fig.~\ref{fig:lavalle_figs}.
We now return to the prediction of the CR antiproton flux, for which the involved physics is a little more complex 
than for gamma-ray calculations. The additional steps come from that charged CRs scatter off magnetic turbulences 
in the Galaxy (which drives their motion to a diffusion pattern), are convected away from the disk by Galactic 
winds, and suffer nuclear reactions (spallation) with the interstellar medium (ISM) as well as energy 
redistribution due to scattering interactions and diffusive reacceleration. For a complete theoretical view on 
CR propagation, see Ref.~\cite{berezinsky_book_90}. Thus, a full prediction relies on solving a diffusion 
equation that reads (steady state and stable species):
\ben
\label{eq:diff_eq}
-\vec \nabla \{  K_x \,\vec \nabla\, {\cal N} - \vec V_c \, {\cal N}  \} + 
\partial_E \left\{  \widehat { \cal K}_E \, {\cal N} \right \} 
+ \Gamma_s \, {\cal N}= { \cal Q}(\vec x,E)\;.
\een
In this equation, ${\cal N} = dn/dE$ is the CR density per unit energy, $K_x$ is the spatial diffusion 
coefficient, $V_c$ is the convective wind velocity, $ \Gamma_s$ is the rate of spallation with the ISM gas and $\widehat 
{ \cal K}_E $ is an energy redistribution (differential) operator. The rhs term is the source term. For DM 
annihilation, it is proportional to the product of the annihilation cross section and the squared 
DM number density 
(the squared DM mass density to the squared WIMP mass), $\sigv \, (\rho/\mchi)^2$. Semi-analytic solutions to 
the previous equation do exist in certain regimes, especially when one fixes the geometry of the diffusion zone 
to a slab inside which the diffusion coefficient is set homogeneous. Further reasonably fixing the spatial 
extent of this slab, \ie~its radius $R\approx 20$ kpc and half-width $L \approx 4$ kpc, closes the full 
definition of the propagation model. All these parameters are globally referred to as transport parameters. They
are usually constrained with CR nuclei data --- basically, one predicts the fluxes of those secondary 
species arising from the spallation of some primary astrophycical CR nuclei, the source of which 
may be rather well 
modeled (the {\em a priori} unknown source parameters cancel out at zeroth order when using the 
secondary-to-primary ratios --- \eg~\cite{2001ApJ...555..585M}).
Predictions of the CR antiproton flux due to DM annihilation have been derived for years
(\eg~\cite{2005PhRvD..72h3518B}), but were often inconclusive because of uncertainties in the diffusion halo size,
a diffusion zone down to $L\approx 1$ kpc inducing an unobservable flux. Nevertheless, the constraints on the 
diffusion volume have been strongly improved since then, and it is now expected to extend 
to $L\gtrsim 3$ kpc \cite{2010A&A...516A..66P} --- it is also interesting to note that taking $L\approx 1$ kpc 
inevitably leads to a strong excess of secondary astrophysical positrons below 5 GeV 
\cite{2009A&A...501..821D,2011MNRAS.414..985L}. The results to be presented below were essentially 
derived using $L=4$ kpc, and a diffusion coefficient normalization of $K_0 = K_x(p/q = 1 GV)\simeq 0.01 \,
{\rm kpc^2/Myr}$.

Predictions of the DM-induced antiproton flux are poorly sensitive to the distribution of DM close to the 
Galactic center --- as long as $\rho \propto r^{-\gamma}$ with $\gamma\lesssim 1.5$ ---  in contrast to 
gamma-ray predictions. This is due to diffusion itself, that tends to smear out far away inhomogeneities. Actually,
the antiproton flux is much better controlled by the local DM density $\rho_\odot$. There has been some recent 
attempts to bracket it, and it is noteworthy that quite different methods have led to similar 
results, $\rho_\odot \simeq 0.4$, with a $\sim 20\%$ accuracy (\eg~
\cite{2010JCAP...08..004C,2010A&A...523A..83S,2011MNRAS.tmp..553M}). The coming results can be derived 
by using the best-fit Einasto \cite{1969Afz.....5..137E} or Navarro-Frenk-White \cite{1997ApJ...490..493N} 
profiles provided in Refs. \cite{2010JCAP...08..004C} and \cite{2011MNRAS.tmp..553M}, indifferently.

As to DM models, we are going to consider two template cases that are likely representative of almost
all quarkophilic configurations. In the first one~\cite{2010PhRvD..82h1302L}, annihilation fully goes 
into $b\bar b$-quark pairs, which is conservatively representative of annihilation into any quark pair, up to 
an antiproton multiplicity factor typically $\lesssim 2$  in the interesting $\sim 0.1-5$ GeV kinetic 
energy range, unfavorable to heavy quarks (smaller dynamical range). In the second one~\cite{2012NuPhB.854..738C}, 
annihilation goes into two unstable particles, scalar and pseudo-scalar, with any masses above 1 GeV
allowed by kinematics, and which may further decay into quarks. These (pseudo-)scalar particles could be thought of as 
the light Higgs bosons arising in the NMSSM, or more generally in any singlet extension of the MSSM. Except for 
spin considerations, this annihilation channel is kinematically representative of any two-body intermediate state 
made of unstable bosons. The following analysis can therefore be extrapolated, for instance, to models that 
involve couplings to new light gauge fields~\cite{2011arXiv1108.1391C}.

The antiproton constraints merely come from the fact that the expected secondary astrophysical background
already saturates the existing data, \eg~from BESS~\cite{2011arXiv1107.6000A} or PAMELA~\cite{2010PhRvL.105l1101A}. 
Consequently, any other significant astrophysical yield quickly leads to an excess prediction. The important point 
is that the theoretical errors affecting the astrophysical background estimates are small, at the 15\% level 
below a few GeV~\cite{2001ApJ...563..172D}. It turns out that the DM-induced antiproton flux generically exceeds 
the background expectation for WIMP masses between $\sim$5-15 GeV and $\sigv \geq 10^{-26}\,{\rm cm^3/s}$, for the 
two configurations discussed above. Therefore, if the annihilation cross setion is S-wave dominated at the decoupling 
time in the early universe, then requiring the correct DM relic abundance, which translates into 
having $2.5\times 10^{-26} \lesssim \sigv \lesssim 5\times 10^{-26}\,{\rm cm^3/s}$, generically leads to an antiproton 
excess in this mass range --- a smaller cross section is difficult to achieve, since leading to an over-abundance of 
DM. Examples are given in the middle and right panels of Fig.~\ref{fig:lavalle_figs}. The only ways left to escape 
these antiproton bounds are (i) to consider the second configuration, \ie~the annihilation into a 
two-boson intermediate state where the decaying bosons would both have masses lower than the (anti)proton mass, 
or (ii) to be at the $b$-quark pair threshold production at which antiprotons can also barely be produced.
The first case might be easy to cook up, but the second one is clearly very fine-tuned.
\section{Conclusion}
\label{sec:concl}
We have discussed how CR antiprotons can put limits on light DM candidates. They are 
generically strong and exclude quarkophilic DM particles in the 5-15 GeV mass range relevant to the 
direct detection signals, if they annihilate through S-waves with $\sigv \geq 10^{-26}\,{\rm cm^3/s}$. The only 
constraint at the origin of this result is the upper bound on the cosmological DM abundance, which translates into 
a lower limit on the annihilation cross section at the decoupling time, $\sigv \gtrsim 2.5\times 10^{-26}
\,{\rm cm^3/s}$. Light neutralinos arising in the MSSM with non-unified gaugino masses and in the NMSSM are therefore 
excluded as potential DM candidates if they fall in the quarkophilic categories discussed above, which is quite often 
the case. These antiproton limits are complementary to other studies using \eg~gamma-rays 
\cite{2011arXiv1108.2914G}, CMB photons \cite{2011PhRvD..84b7302G}, and solar neutrinos \cite{2011NuPhB.850..505K}, 
which concern more annihilation final states than only quarks.
\bibliographystyle{iopart-num}
\bibliography{lavalle_bib}
\end{document}